\documentclass[runningheads]{llncs}

\usepackage{lineno,hyperref}
\modulolinenumbers[5]

\usepackage{amsmath,amssymb}
\usepackage{balance}
\usepackage{color}
\usepackage{graphicx}
\usepackage{multirow}
\usepackage{url}
\usepackage{array}
\usepackage{url}
\usepackage[width=122mm,left=12mm,paperwidth=146mm,height=193mm,top=12mm,paperheight=217mm]{geometry}

\urldef{\mail}\path|zhengyb@sari.ac.cn|
\newcommand{\keywords}[1]{\par\addvspace\baselineskip
\noindent\keywordname\enspace\ignorespaces#1}

\newcommand\ie{{i.e.}}

\newcommand\eg{{e.g.}}

\begin{document}

\mainmatter

\title{Satellite Image Scene Classification via\\ ConvNet with Context Aggregation}

\titlerunning{Satellite Image Scene Classification via ConvNet with Context Aggregation}

\author{Zhao Zhou\inst{1} \and
        Yingbin Zheng\inst{1}\thanks{Corresponding author.} \and
        Hao Ye\inst{1} \and
        Jian Pu\inst{2} \and
        Gufei Sun\inst{3}}

\authorrunning{Z. Zhou et al.}

\institute{Shanghai Advanced Research Institute, Chinese Academy of Sciences \and
           East China Normal University, Shanghai, China \and
           ZhongAn Technology, Shanghai, China\\
           \mail}

\maketitle

\begin{abstract}

Scene classification is a fundamental problem to understand the high-resolution remote sensing imagery. Recently, convolutional neural network (ConvNet) has achieved remarkable performance in different tasks, and significant efforts have been made to develop various representations for satellite image scene classification. In this paper, we present a novel representation based on a ConvNet with context aggregation. The proposed two-pathway ResNet (ResNet-TP) architecture adopts the ResNet \cite{2016cvpr_khe} as backbone, and the two pathways allow the network to model both local details and regional context. The ResNet-TP based representation is generated by global average pooling on the last convolutional layers from both pathways. Experiments on two scene classification datasets, UCM Land Use and NWPU-RESISC45, show that the proposed mechanism achieves promising improvements over state-of-the-art methods.

\end{abstract}

\section{Introduction}
\label{sec:intro}

With the growing deployment of remote sensing instruments, satellite image scene classification, or scene classification from high-resolution remote sensing imagery, has drawn attention for its potential applications in various problems such as environmental monitoring and agriculture. Multiple challenges exist to produce accurate scene classification results. Large intra-class variation in the same scene class is a common issue. Moreover, the semantic gap between the scene semantic and the image features could further increase the difficulties of robust classification. Thus, the design of suitable representations on satellite images to deal with the challenges is of fundamental importance.

Great progress has been achieved in the recent years with the utility of representations based on the convolutional neural networks (ConvNet), which led to   breakthroughs in a number of computer vision problems like image classification.
The typical ConvNet including AlexNet~\cite{2012nips_aKrizhevsky}, SPP-net~\cite{2015tpami_khe}, VGG~\cite{2015ICLR_kSimonyan}, and GoogleNet~\cite{2015cvpr_cszegedy}, has also been applied to the task of satellite image scene classification.
As the image number of the satellite image datasets are order-of-magnitude smaller than that of the image classification datasets (\eg, ImageNet~\cite{2009cvpr_jdeng}) and may not sufficient to train the robust deep models, these ConvNet based methods usually employ the off-the-shelf pre-trained deep networks (\eg, in  \cite{2017TGRS_Qliu,2017GRSL_GScott,Han2017Pre,Xia2016AID,2017rs_xhan,2017pieee_gcheng,2017GRSL_GCheng,2017JSTARS_GWang,8086123}). The activations of the layers or their fusion are considered as the visual representation and sent to the scene classifiers. Evaluations on the benchmarks show that the deep learning based features often outperform previous handcrafted features.

The number of stacked layers in most current deep networks for satellite images is relatively small. For example, \cite{2017rs_xhan} design classification systems based on the 7-layer  architecture of AlexNet~\cite{2012nips_aKrizhevsky} or its replication CaffeNet~\cite{2014mm_yjia}, and \cite{2017pieee_gcheng,2017GRSL_GCheng} employ the 16-layer VGG architecture~\cite{2015ICLR_kSimonyan}. Recent evidence suggests that deeper convolutional networks are more flexible and powerful with high modeling capacity for image classification~\cite{2016cvpr_khe,huang2017densely}. Some previous works (\eg, \cite{2017JSTARS_GWang,2017GRSL_GScott}) employ the Residual Networks (ResNet)~\cite{2016cvpr_khe} as one of the basic models. However, the effectiveness of these deeper models and how their performance depends on the number of layers are still not fully exploited for remote sensing images.

\begin{figure*}[t]
  \centering
  \includegraphics[width=.8\linewidth]{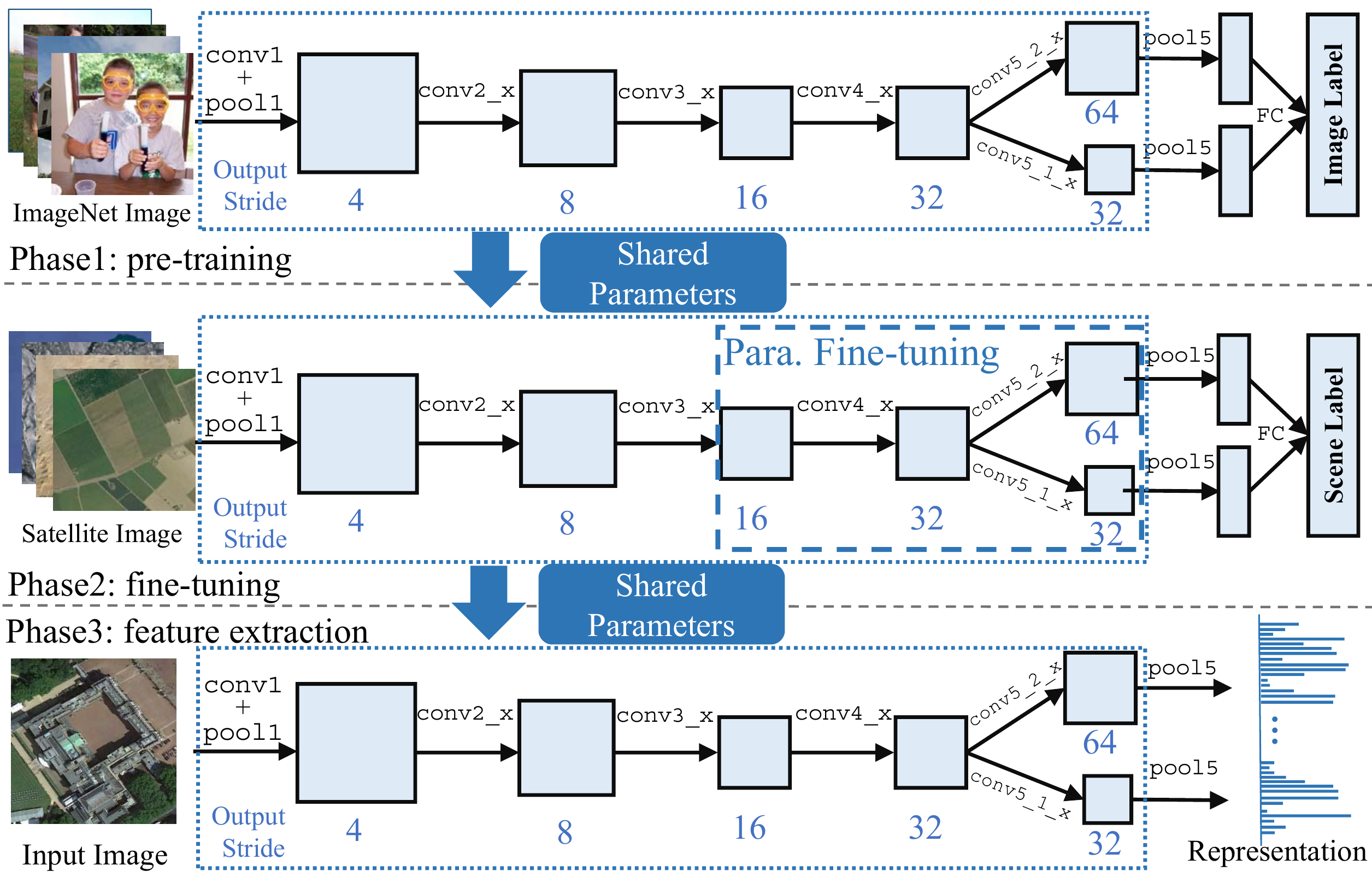}
  \caption{Pipeline of the proposed framework with two-pathway ResNet (ResNet-TP). The network is pre-trained using ImageNet database (Phase 1). Phase 2 produces the fine-tuned network with the satellite image dataset. A given satellite image goes through the network and the representation is generated from the global average pooling on the last convolutional layers (Phase 3).
  }
  \label{fig:pipeline}
\end{figure*}

In this work, we focus on the problem of deeper ConvNet with context aggregation, and introduce an image representation built upon a novel architecture for satellite images, which adopts the ResNet~\cite{2016cvpr_khe} as backbone. The two-pathway ResNet (or ResNet-TP abbreviatedly) is proposed, and Fig. \ref{fig:pipeline} illustrates the pipeline. The proposed structure aims to aggregate the contextual information to enhance the feature discrimination. The input images go through two paths of convolutional operations after a few layers: one path  follows the default building block settings, and another path incorporates the dilation within convolutional layers to expand the receptive field.
Training the deeper ConvNet is usually more difficult and may lead to higher risk of overfitting, especially when using the relatively small remote sensing dataset. Therefore, we also employ the transfer learning strategy to reuse the parameters learned from image classification dataset. The idea of constructing contextual representations has been taken in several previous remote sensing works, e.g., \cite{2017rs_xhan} and \cite{2017TGRS_Qliu}. These approaches use a single spatial pyramid pooling~\cite{2015tpami_khe} on the feature maps of last convolutional layer, which is usually tiny after progressively resolution reduction of previous operations. ResNet-TP is designed with contextual pathways \emph{before} last convolutional layers, and is able to alleviate the loss of spatial acuity caused by tiny feature maps. To evaluate our proposed framework, we report the evaluations on the recent NWPU-RESISC45~\cite{2017pieee_gcheng} and the UC Merced (UCM) Land Use dataset \cite{yang2010bag}. Our representation is compared with several recent approaches and achieves state-of-the-art performance.

\section{Methodology}
\label{sec:dcnn}

\subsection{ResNet Architecture}
\label{sec:resnet}

\begin{figure}[t]
  \centering
  \includegraphics[width=.5\linewidth]{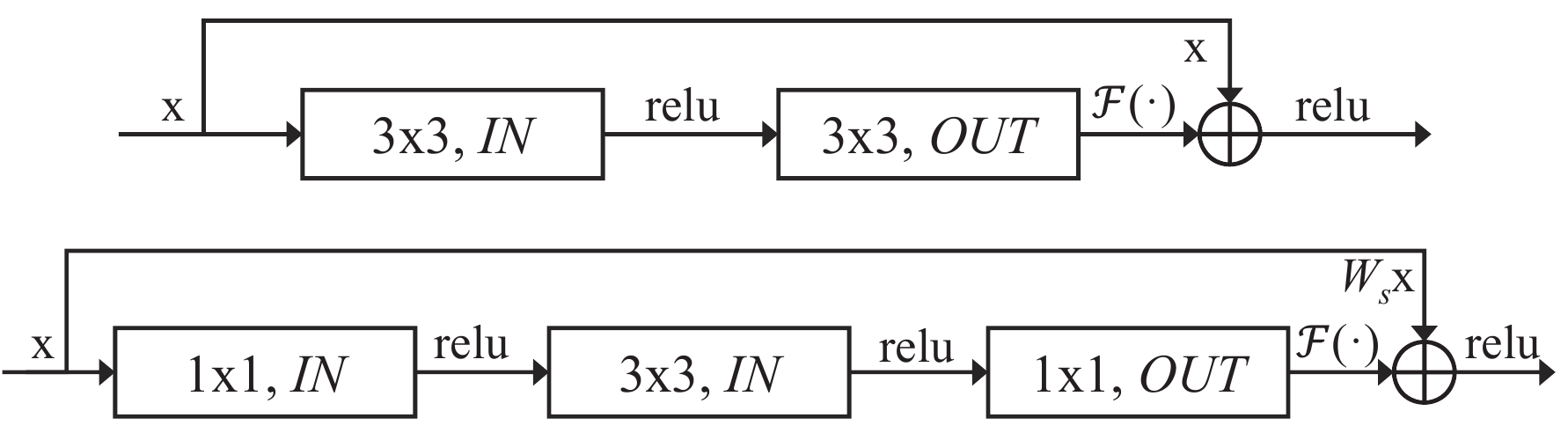}
  \caption{The building block with the residual function $\mathcal{F}$. $IN$ and $OUT$ denote number of in-plane and out-plane, respectively. Top: the basic building block. Bottom: the bottleneck building block.}
  \label{fig:residual}
\end{figure}

We begin with a brief review of the ResNet and the residual learning to address the training issue of deeper neural networks, which is the foundation to win the ILSVRC\&COCO 2015 competition for the tasks of ImageNet detection, ImageNet localization, COCO detection, and COCO segmentation \cite{2016cvpr_khe}. First, downsampling is performed directly by one $7\times7$ convolutional layer and one max-pooling (over $2\times2$ pixel window) with stride 2, respectively. The main component used to construct the architecture is the stacked convolutional layers with shortcut connections. Such building block is defined as
\begin{equation}
    \mathcal{H}(\mathbf{x}) = \mathcal{F}(\mathbf{x},\{W_i\})+W_s\mathbf{x},
\end{equation}
where $\mathbf{x}$ and $\mathcal{H}(\mathbf{x})$ are the input and output of the building block, $\mathcal{F}(\cdot)$ is the residual mapping function to be learned, $W_i$ is the parameters of the convolutional layers, and $W_s$ is linear projection matrix to ensure the dimension matching of $\mathbf{x}$ and $\mathcal{F}$ ($W_s$ is set as identity matrix when they are with the same dimension). The operation $\mathcal{F}(\cdot)+W_s\mathbf{x}$ is performed by a shortcut connection and element-wise addition. There are usually two or three layers within one building block, and two typical building blocks are shown in Fig. \ref{fig:residual}, where the basic building block is for 18/34-layer and the bottleneck building block is for 50/101/152-layer in \cite{2016cvpr_khe}. The convolution is performed with the stride of 2 after a few building blocks to reduce the resolution of feature maps. Unlike previous architectures such as AlexNet~\cite{2012nips_aKrizhevsky} and VGG~\cite{2015ICLR_kSimonyan}, ResNet has no hidden fully-connected (FC) layers; it ends with a global average pooling and then a $N$-way FC layer with softmax ($N$ is the number of classes). We refer the reader to \cite{2016cvpr_khe} for more details.

\begin{table}[b]
\begin{center}
\caption{Configuration of the groups in ResNet-TP Architecture. Suppose the input image is with size $224\times224$. Basic($IN,OUT$) and Bottleneck($IN,OUT$) denote the basic and bottleneck building block with number of in-plane $IN$ and out-plane $OUT$ (see Fig. \ref{fig:residual}). `$\times n_i$' indicates stacking $n_i$ blocks, where $[n_2,n_3,n_4,n_5]$=[2,2,2,2] for 18 layer, [3,4,6,3] for 34/50 layer, [3,4,23,3] for 101 layer.}
\label{tab:arch}
\begin{tabular}{c|p{3.7cm}<{\centering}|p{3.7cm}<{\centering}|c}
\hline
\multirow{2}{*}{Group} & \multicolumn{2}{c|}{Block} & Output size, \\\cline{2-3}
 & 18/34 layer & 50/101 layer & dilation\\
\hline
{\tt conv1+pool1} &\multicolumn{2}{c|}{[7$\times$7, 64]; Max Pooing} & 56$\times$56, 1\\
\hline
{\tt conv2\_x} & Basic(64,64)$\times n_2$& Bottleneck(64,256)$\times n_2$& 56$\times$56, 1\\
\hline
{\tt conv3\_x} & Basic(128,128)$\times n_3$ & Bottleneck(128,512)$\times n_3$& 28$\times$28, 1 \\
\hline
{\tt conv4\_x}  & Basic(256,256)$\times n_4$ & Bottleneck(256,1024)$\times n_4$& 14$\times$14, 1 \\
\hline
{\tt conv5\_2\_x}  & \multirow{2}{*}{Basic(512,512)$\times n_5$} & \multirow{2}{*}{Bottleneck(512,2048)$\times n_5$}& 14$\times$14, 2\\
{\tt conv5\_1\_x} &   &  & 7$\times$7, 1\\
\hline
\end{tabular}
\end{center}
\end{table}

\subsection{Context Aggregation}
\label{sec:eca}

We now elaborate the construction of ResNet-TP. The architecture of the network is summarized in Table \ref{tab:arch}. In general, the network contains six groups of layers or building blocks. Group {\tt conv1+pool1} consist of the $7\times7$ convolutional layer and the max-pooling, and {\tt conv2\_x} to {\tt conv4\_x} are with a stack of building blocks. All their configurations follow the generic design presented as in Sect. \ref{sec:resnet}, and differ only in the depth of blocks. Consider an input image with $224\times224$ pixels, group {\tt conv4\_x} is with output stride of 16 and thus its feature map size is $14\times14$.

We introduce group with dilation convolutional layers, which has been shown to be effective in many tasks such as semantic segmentation~\cite{2016ICLR_fyu,2016arxiv_LChen}, video analysis \cite{rene2017temporal,xu2018dense}, RGB-D \cite{zheng2018learning}, and DNA modeling \cite{gupta2017dilated}. The two-pathway architecture is made of two streams: a pathway with normal building blocks ({\tt conv5\_1\_x}) and another with larger receptive fields ({\tt conv5\_2\_x}). The dilation is operated on the $3\times3$ convolutional layer in the building block. Let $\mathbf{x}$ be the input feature map and $\mathbf{w}$ be the filter weights associated with the dilation convolutional layer, the output $\mathbf{y}$ for position $\mathbf{p}=(p_1,p_2)$ is defined as:
\begin{equation}
    \mathbf{y}(\mathbf{p})=\sum_{\mathbf{d}\in \mathcal{G}_d}\mathbf{w}(\mathbf{d})\cdot\mathbf{x}(\mathbf{p}+\mathbf{d})
\end{equation}
where $\mathcal{G}_d=\{(-d,-d),(-d,0),\dots,(0,d),(d,d)\}$ is the grid for the $3\times3$ filters and $d$ is the dilation. We set the dilation $d=2$ for {\tt conv5\_2\_x}, and the layers in {\tt conv5\_1\_x} can also be considered as a special case with $d = 1$.
The motivation for this architectural design is that we would like the prediction to be influenced by two aspects: the visual details of the region around each pixel of the feature map as well as its larger context. In fact, ResNet-TP is degenerated to the standard ResNet when {\tt conv5\_2\_x} and its subsequent layers are removed. Finally, we connect the last convolutional hidden layers in both pathways with the global average pooling followed by the FC layer with softmax to perform a prediction of the labels.

\subsection{Model Training and Implementation Details}
\label{sec:train}

The ResNet-TP architecture is with a large amount of parameters to train. A traditional remote sensing dataset contains thousands of high-resolution satellite images, which is far less than the image classification datasets for training the state-of-the-art deep learning models.
Following previous works \cite{2017TGRS_Qliu,2017GRSL_GScott}, training of ResNet-TP is based on the transfer learning strategy and Fig. \ref{fig:pipeline} illustrates the overall framework of the proposed ResNet-TP based scene representation.

The whole training procedure as well as the feature extraction are carried out via the open source PyTorch library and an Nvidia Titan X (Pascal) GPU. The first phase is to get a pre-trained model using the ImageNet database~\cite{2009cvpr_jdeng}. During this process, due to the network with only {\tt conv5\_1\_x} pathway having the same structure with the original ResNet, we set the weights of {\tt conv1\_x} to {\tt conv5\_1\_x} with the existing PyTorch ResNet models\footnote{The download link can be found from \url{https://github.com/pytorch/vision/blob/master/torchvision/models/resnet.py}}. Directly updating model from this initialization lead to performance drop, as the parameters of {\tt conv5\_2\_x} are randomly initialized. On the other hand, it is time-consuming if the model is trained from scratch, since ImageNet contains millions of images. Here we make a compromise by learning the weights of {\tt conv5\_2\_x} and its subsequent layers from the network with only {\tt conv5\_2\_x} pathway and by frozen of {\tt conv1\_x} to {\tt conv4\_x}\footnote{The stochastic gradient descent (SGD) is used with batch size of 64 and momentum of 0.9. The learning rate is initially set to be 0.01 and is divided by 10 every 30 epochs.}. We compare this pre-training strategy with the model trained from scratch under ResNet-TP-18, and find that they are with similar performance on ImageNet validation set, while its training is much faster.

For the fine-tuning phase, we only fine-tune the building block groups after {\tt conv3\_x} by using the training satellite images and their labels due to the limitation of the GPU memory. We take random rotated, mirrored, or scaled images for data augmentation during fine-tuning. Finally, the representation is obtained from the global average pooling in both pathways, and the linear SVM classifier with default setting $C=1$ is carried out for a fair comparison with previous works \cite{2017TGRS_Qliu,2017pieee_gcheng,2017GRSL_GCheng,2017JSTARS_GWang}.


\begin{figure}[t]
  \centering
  \includegraphics[width=\linewidth]{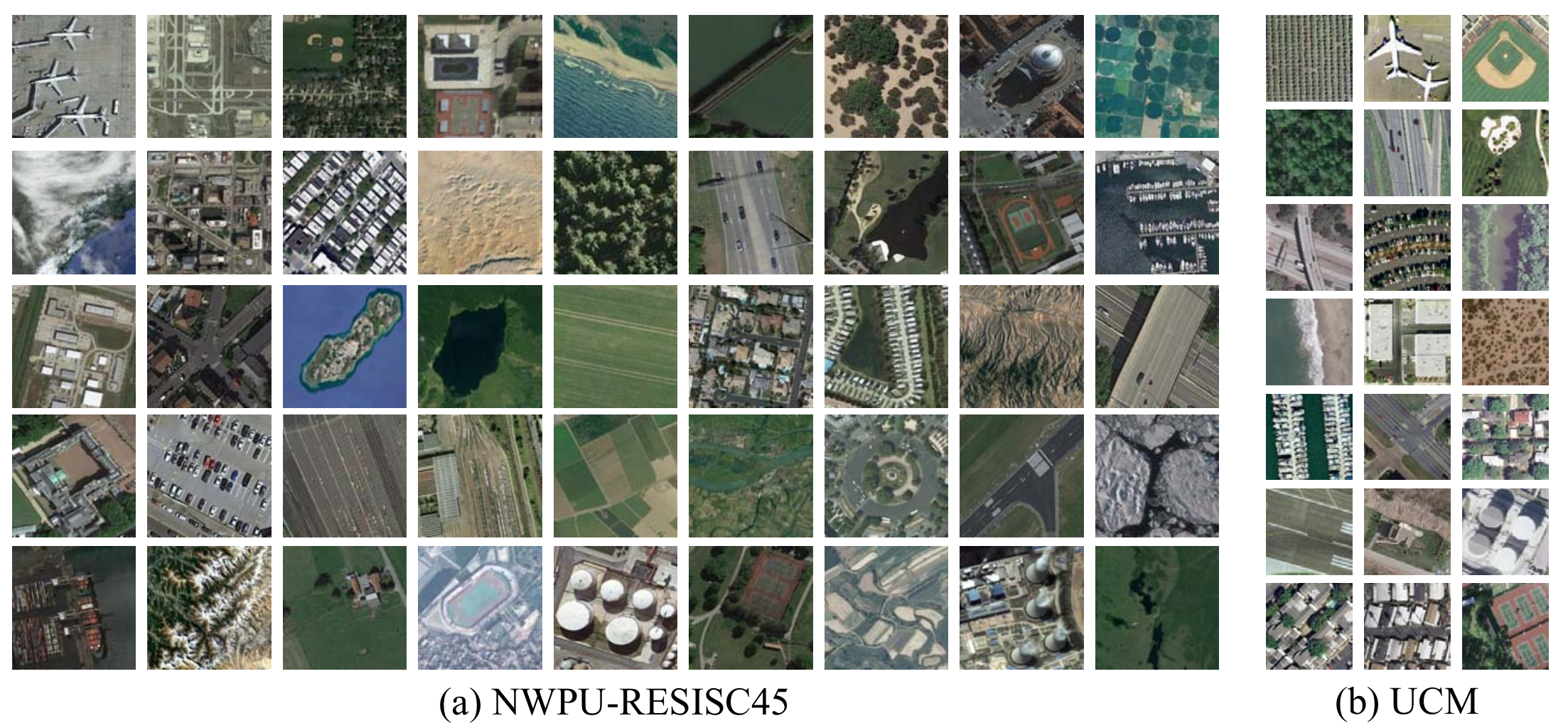}
  \caption{Scene categories from the datasets.}
  \label{fig:datasets}
\end{figure}


\begin{table*}[b]
\begin{center}
\caption{Overall accuracies and standard deviations (\%) of the proposed methods and state-of-the-arts under different training ratios on the NWPU-RESISC45 dataset. The results of pre-trained (PT-*) and fine-tuned (FT-*) ConvNets are reported in \cite{2017pieee_gcheng}, and results of BoCF are from \cite{2017GRSL_GCheng}.}
\label{tab:nwpu}
{\scriptsize
\begin{tabular}{c|p{1.8cm}<{\centering}p{1.8cm}<{\centering}||c|p{1.8cm}<{\centering}p{1.8cm}<{\centering}}
\hline
\multirow{2}{*}{Network} &\multicolumn{2}{c||}{Training ratios}&\multirow{2}{*}{Network} &\multicolumn{2}{c}{Training ratios}\\\cline{2-3}\cline{5-6}
  &10\% & 20\% & &10\% & 20\%\\\hline
 PT-AlexNet  & $76.69 \pm 0.21$& $79.85 \pm 0.13$&BoCF-AlexNet & $55.22\pm 0.39$& $59.22 \pm 0.18$\\
 PT-GoogleNet  &$76.19\pm 0.38$ &$78.48\pm 0.26$ &BoCF-GoogleNet &$78.92\pm 0.17$ &$80.97\pm 0.17$\\
PT-VGG-16 &$76.47\pm 0.18$ &$79.79\pm0.15$ &BoCF-VGG-16 &$82.65\pm 0.31$ &$84.32\pm0.17$ \\\hline
 FT-AlexNet  & $81.22 \pm 0.19$& $85.16 \pm 0.18$&Triplet networks\cite{8086123} &- &$92.33\pm0.20$\\\cline{4-6}
 FT-GoogleNet  &$82.57\pm 0.12$ &$86.02\pm0.18$ &ResNet-TP-18 &$87.79\pm 0.28$ &$91.03\pm0.26$\\
FT-VGG-16 & $87.15\pm 0.45$ &$90.36\pm 0.18$ &ResNet-TP-101 &{\bf 90.70$\pm$ 0.18} &{\bf 93.47$\pm$0.26}\\\hline
\end{tabular}
}
\end{center}
\end{table*}

\section{Experiments}
\label{sec:evaluation}

To evaluate the effectiveness of the proposed method, we compare it with several state-of-the-art approaches on two remote sensing scene classification datasets, including the recent proposed 45-Class NWPU-RESISC45 dataset \cite{2017pieee_gcheng} and the widely used 21-Class UCM Land Use dataset \cite{yang2010bag}.

\subsection{NWPU-RESISC45}

The NWPU-RESISC45 dataset contains 31500 remote sensing images extracted from Google Earth covering more than 100 countries and regions. Each scene class is composed of 700 images with the spatial resolution varied from about 30 to 0.2 m per pixel. Sample images and the scene categories are shown in Fig. \ref{fig:datasets}(a), and we wrap the images into the size of $224\times224$. We follow the official train/test split strategy with two training ratios, \ie, 10\% (10\% for training and 90\% for testing) and 20\% (20\% for training and 80\% for testing). We repeat the evaluations ten times under each training ratio by randomly splitting the dataset and also report the mean accuracy and standard deviation.

\begin{figure*}[t]
  \centering
  \includegraphics[width=.9\linewidth]{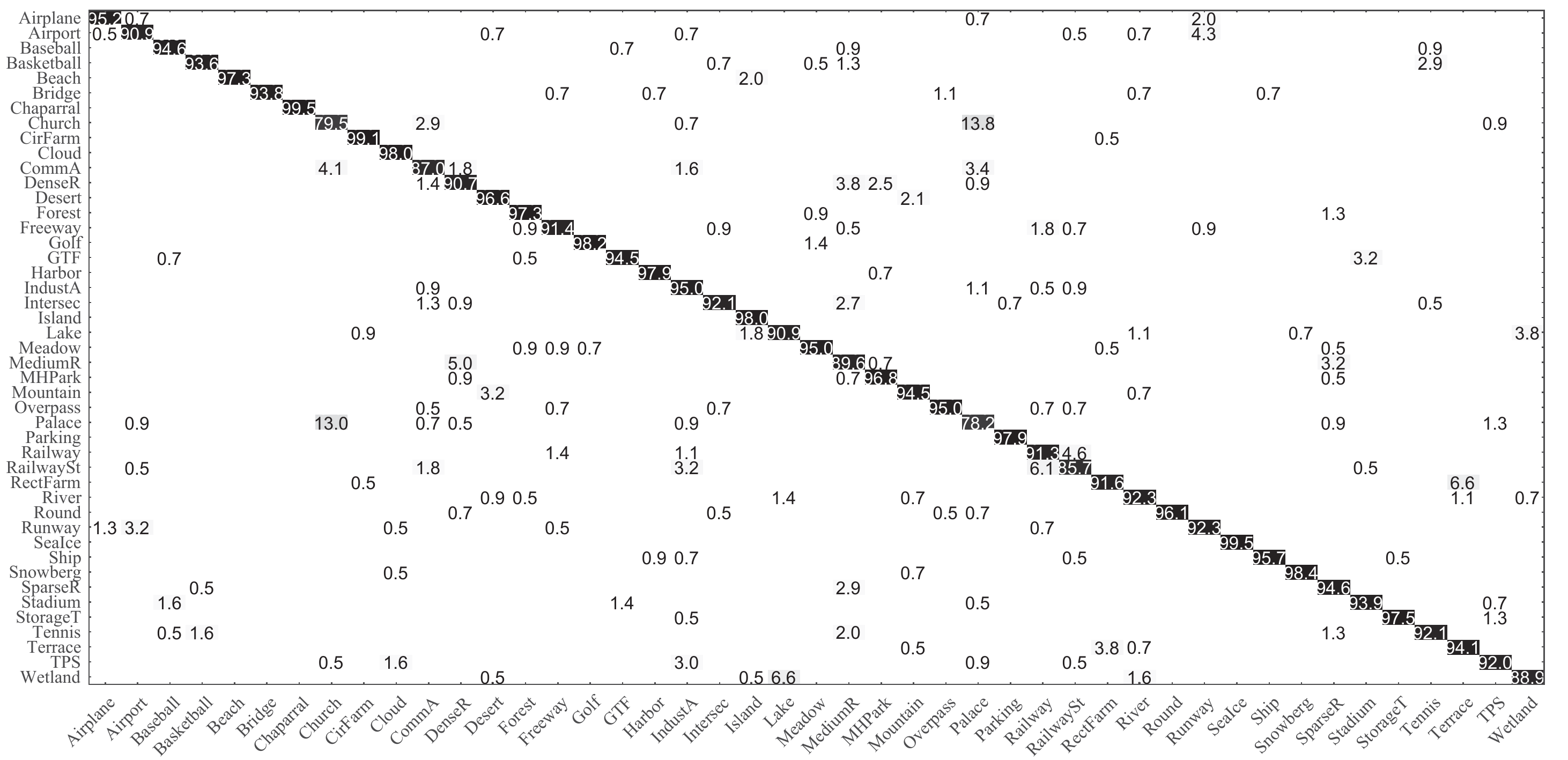}
  \caption{Confusion matrices under the training ratio of 20\% by using ResNet-TP-101 on the NWPU-RESISC45 dataset.}
  \label{fig:nwpu_cm}
\end{figure*}

\begin{figure}[t]
  \centering
  \includegraphics[width=.82\linewidth]{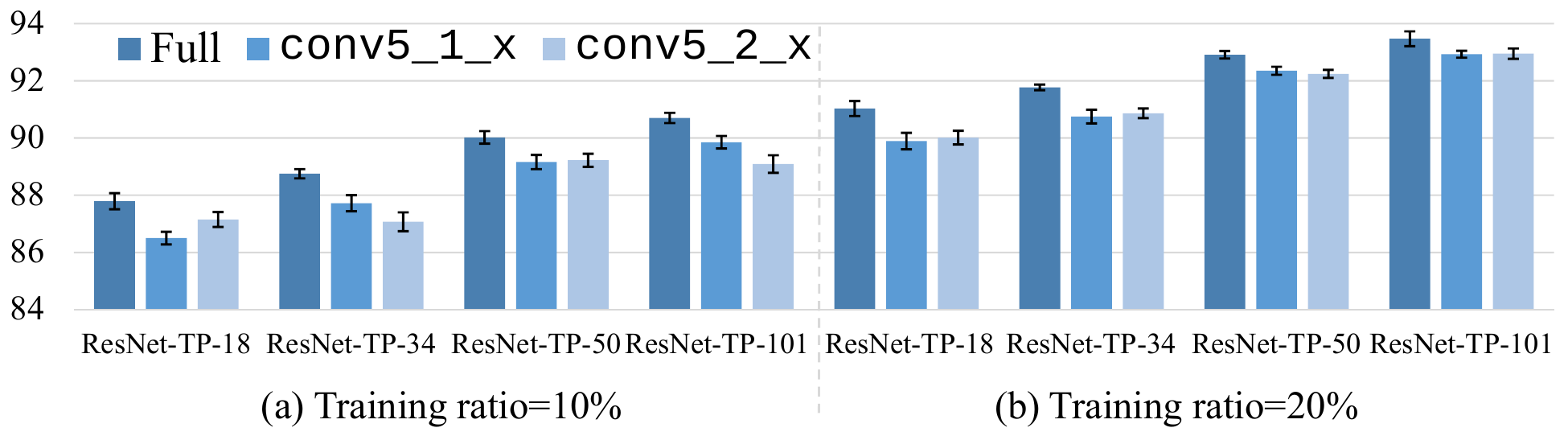}
  \caption{Evaluation of the ResNet-TP parameters and components with with different training ratio on the NWPU-RESISC45 dataset. {\tt conv5\_1\_x} and {\tt conv5\_2\_x} indicates the network with only one stream
of ResNet-TP.}
  \label{fig:nwpu_para}
\end{figure}

\begin{figure}[t]
  \centering
  \includegraphics[width=.7\linewidth]{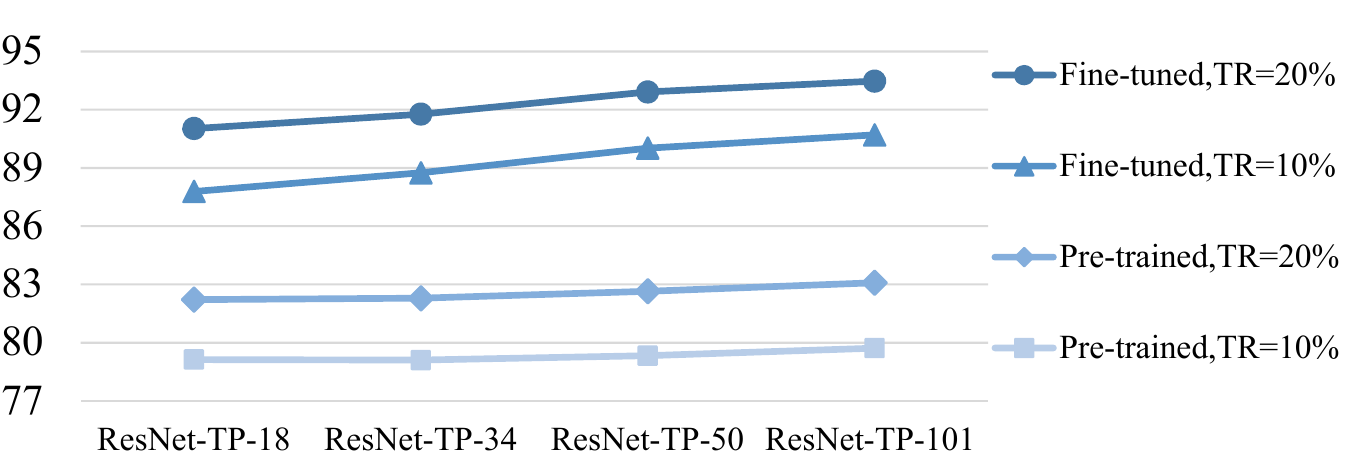}
  \caption{Comparison of the pre-trained and fine-tuned ResNet-TP models on the NWPU-RESISC45 dataset. TR indicates training ratio.}
  \label{fig:nwpu_pt}
\end{figure}

We compare ResNet-TP based representation with several baselines and state-of-the-art approaches. Among them, the first group contains several well-known baseline descriptors, including the pre-trained or fine-tuned AlexNet \cite{2012nips_aKrizhevsky}, GoogleNet \cite{2015cvpr_cszegedy}, and VGG-16~\cite{2015ICLR_kSimonyan}. Table \ref{tab:nwpu} shows that the proposed representation outperforms all the baseline descriptors as well as the state-of-the-art approaches shown in the right part of Table \ref{tab:nwpu}, including the Bag of Convolutional Features (BoCF) \cite{2017GRSL_GCheng} and a very recent triplet networks~\cite{8086123}. In Fig. \ref{fig:nwpu_cm}, we report the confusion matrix and detail classification accuracy for each scene label using training ratios of 20\%.

\emph{Network and Training Ratio.} We also study the performance of different network settings and training ratios, and results are given in Fig. \ref{fig:nwpu_para}. Adding the layers in ResNet-TP architecture and context aggregation by the two-pathways boost the classification accuracy.
We conjecture that the applied single pathway may not be the one at which the network responds with optimal
confidence, and context aggregation with multiple pathways increase the robustness.
In addition, we observe that increasing training images (70 to 140 per class) lead to significant performance gains, probably due to the scene variation and data diversity in the NWPU-RESISC45 dataset.

\emph{Pre-Trained }vs. \emph{Fine-Tuned.} Our last experiment on NWPU-RESISC45 evaluates the alternative method for ResNet-TP model generation. While the fine-tuned method follows the pipeline of Fig. \ref{fig:pipeline}, the pre-trained approach is only composed of phase 1 and 3 in the figure, and the model parameters are directly learned from ImageNet. The comparison between the curves in Fig. \ref{fig:nwpu_pt} verifies that for both training ratios using the fine-tuned network is important toward a more discriminant representation. We also find that the fine-tuned method outperforms the pre-trained method even though the training images is half of which for pre-trained.

\subsection{UCM Land Use}

\begin{figure}[t]
  \centering
  \includegraphics[width=.9\linewidth]{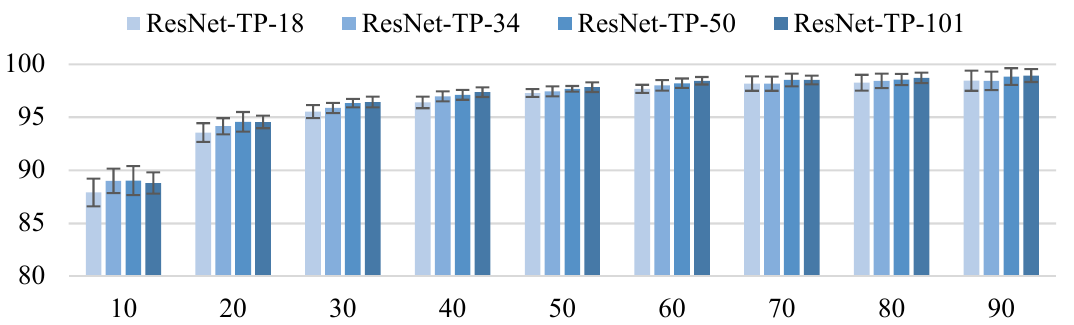}
  \caption{Evaluation of the ResNet-TP models with different number of training images on the UCM dataset.}
  \label{fig:ucm_number}
\end{figure}

The UCM Land Use dataset contains 2100 aerial scene images extracted from United States Geological Survey (USGS) national maps. Each land use class is composed of 100 images with the spatial resolution of 1 ft and the size of $256\times256$ pixels. The sample images are illustrated in Fig. \ref{fig:datasets}(b).
As UCM Land Use dataset is with relatively small and the results on it are already saturated, in this paper we focus on the performance w.r.t. the number of training images.
Fig. \ref{fig:ucm_number} shows the effect of training image number in the representation. We observe significant performance gains when the number of training images increases from 10 to 50, after which the performance tends to be saturated. Another observation is that the result of ResNet-TP-50 is similar to the accuracy of ResNet-TP-101 in most of the comparisons, indicating that the computation could be saved by ResNet-TP-50 with marginal performance drop.

\begin{table}[t]
\begin{center}
\caption{Overall accuracies and standard deviations (\%) of the proposed methods and state-of-the-arts on the UCM dataset. `-' indicates that the results are not available in the corresponding paper.}
\label{tab:ucml}
\begin{tabular}{c|p{2.2cm}<{\centering}p{2.2cm}<{\centering}p{2.2cm}<{\centering}}
\hline
Number of images & 5 & 50&80\\\hline
MKL \cite{Cusano2015Remote} & $64.78\pm1.62$&$88.68\pm1.10$ &$91.26\pm1.17$ \\\hline
SPP-net MKL \cite{2017TGRS_Qliu} & $75.33\pm1.86$&$95.72\pm0.50$ &$96.38\pm0.92$ \\\hline
AlexNet-SPP-SS \cite{Han2017Pre} & - & - & $96.67\pm0.94$\\\hline
VGG-16 \cite{Xia2016AID} & - & $94.14\pm0.69$ & $95.21\pm1.20$\\\hline
ResNet50 \cite{2017GRSL_GScott}& - & - & $98.50\pm1.40$\\\hline
ResNet-TP-50 & {\bf 77.07$\pm$1.73} & {\bf 97.68$\pm$0.26} & {\bf 98.56$\pm$0.53}\\\hline
\end{tabular}
\end{center}
\vspace{-0.2in}
\end{table}

We also compare the results of proposed representation with several state-of-the-art approaches. Table \ref{tab:ucml} summarizes the overall accuracy and standard deviation of all the classes.
As can be seen from the table, the ResNet-TP based representation shows very competitive performance with different number of training images, which is significantly better than the other representations when the training images are limited. We also notice previous approach ResNet152\_EMR \cite{2017JSTARS_GWang} is also a ResNet-152 based representation and reach the accuracy of 98.90\% by combining information from multiple layers with larger input image size ($320\times320$). When the input image size is set to $224\times224$, the classification accuracy is 98.38\%, which is inferior to ours with fewer layers. We believe that ResNet-TP based representation is also complementary to these mixed-resolution methods since they focus on different levels of information, which will be examined in the future work.

%
%

\section{Conclusion}
\label{sec:conclusion}

In this work, we have introduced ResNet-TP, a two-pathway convolutional network with context aggregation to generate a discriminant representation for satellite image scene classification. Through empirical scene classification experiments, we have shown that proposed ResNet-TP based representation is more effective than previous deep features, generating very competitive results on the UCM Land Use and NWPU-RESISC45 datasets. For future work, we plan to incorporate multi-scale and multiple layers into the ResNet-TP based representation, and also explore the performance benefits of a combination of this representation with other features.

\vspace{0.05in}
\noindent \textbf{Acknowledgments.} This work was supported in part by grants from National Natural Science Foundation of China (No. 61602459) and Science and Technology Commission of Shanghai Municipality (No. 17511101902 and No. 18511103103).

\bibliographystyle{splncs}
\bibliography{total}

\end{document}